\documentclass[aps,prl,groupedaddress,amsmath,amssymb,amsfonts,showpacs,twocolumn]{revtex4}
\usepackage{bm}
\usepackage{graphicx}
\usepackage{color}
\usepackage{psfig}

\begin{document}

%\addtolength{\textheight}{.7cm}

\title{
Energy dependence of current noise in superconducting/normal metal
junctions}
\author{M. Houzet$^1$ and F. Pistolesi$^2$}
\affiliation{
$^1$Commissariat \`{a} l'\'{E}nergie Atomique, DSM,
D\'{e}partement de Recherche Fondamentale sur la Mati\`{e}re
Condens\'{e}e, SPSMS, F-38054 Grenoble, France
\\
$^2$
Laboratoire de Physique et Mod\'elisation des Milieux Condens\'es,\\
CNRS-UJF B.P. 166, F-38042 Grenoble, France
}
\date{\today}

\newcommand{\qav}[1]{\left\langle #1 \right\rangle}
\newcommand{\myT}{\Gamma}
\newcommand{\rem}[1]{}
\newcommand{\FORSE}[1]{{\bf Peut-etre #1} }
\newcommand{\refe}[1]{(\ref{#1})}
\newcommand{\refE}[1]{Eq.~(\ref{#1})}
\newcommand{\beq}{\begin{equation}}
\newcommand{\eeq}{\end{equation}}
\newcommand{\cg}{\check g}
\newcommand{\inc}{{\rm inc}}

\begin{abstract}
Interference of electronic waves undergoing Andreev reflection in
diffusive conductors determines the energy profile of the conductance
on the scale of the Thouless energy.
A similar dependence exists in the current noise, but its
behavior is known only in few limiting cases.
We consider a metallic diffusive wire connected to a
superconducting reservoir through an interface characterized by an
arbitrary distribution of channel transparencies.
Within the quasiclassical theory for current fluctuations we
provide a general expression for the energy dependence of the
current noise.
\end{abstract}

\maketitle

Interference of electronic waves in metallic disordered conductors
is responsible for weak localization corrections to the
conductance \cite{Weakloc}.
If these are neglected,  the probability of transferring an
electron through the diffusive medium is given by the sum of the
{\em modulus squared} of the quantum probability amplitudes for
crossing the sample along all possible paths.
This probability is denoted as semiclassical, since quantum
mechanics is necessary only for establishing the probability for
following each path independently of the phases of the quantum
amplitudes.
In superconducting/normal metal hybrid structures, interference
contributions are not corrections, they may actually {\em
dominate} the above defined semiclassical result for temperatures and
voltages smaller than the superconducting gap.
This is seen experimentally as an energy dependence of the
conductance on the scale of the Thouless energy.
Indeed, the energy dependence comes from the small wavevector
mismatch, linear in the energy of the excitations, between the
electron and the Andreev reflected hole.
This is responsible for the phase difference in the amplitudes for
two different paths leading to interference.
The effect is well known and explicit predictions and measurements
exist for a number of systems \cite{VZK93,Hekking93,Beenakker97}.

Interference strongly affects the current noise too
\cite{BlanterButtiker}.
The largest effects are expected in the tunneling limit, when the
transparency of the barrier is small and its resistance is much
larger than the resistance of the diffusive normal region.
Then, the conductance has a strong non linear dependence at low
bias (reflectionless tunneling) \cite{VZK93,Hekking93}.
This is actually the case, but the zero-temperature noise (or shot
noise) does not give any additional information on the system
since it is simply proportional to the current, as shown
numerically in a specific example in Ref. \cite{SH} and quite
generally in Ref. \cite{PBH}.
The double tunnel barrier system has been considered in Ref. \cite{samuelsson}.
In the case of a diffusive metal wire in contact
with a superconductor through an interface of conductance $G_B$ much
larger than the wire conductance $G_D$, Belzig and Nazarov
\cite{BN01} found that the differential shot noise, $dS/dV$, shows a
reentrant behavior, as a function of the voltage bias, {\em
similar}, but not {\em identical}, to the conductance one.
(The extension of the Boltzman-Langevin approach to the
coherent regime in Ref.\ \cite{houzet} neglects this difference.)
In order to compare quantitatively with actual experiments
\cite{Jehl,KSP,Lefloch} and to gain more insight in the
interference phenomenon, it is necessary to obtain the energy
dependence of noise in more general situations.
The numerical method used in Ref.~\cite{BN01} is, in principle,
suitable to treat more general cases, notably the case when $G_D
\gtrsim G_B $, but only if all channel transparencies,
$\{\myT_n\}$, that characterize the interface are small.
When arbitrary transparencies are present one has to solve
numerically an additional self-consistent equation \cite{private}.
This appears particularly heavy numerically if one is interested to treat a
distribution of transparencies.
We are not aware of results in this direction.
In this paper we present an analytical solution for the
diffusion-type differential equation for the noise within the
theory of current fluctuations \cite{FCS} in the quasiclassical
dirty limit \cite{BN01}.
It allows to treat the general case of arbitrary values for
$\{\myT_n\}$ and $G_B/G_D$.
We express the noise in terms of a function that satisfies a
linear differential equation to be solved numerically once the
channel distribution is given.
We can thus isolate the energy-independent (semiclassical)
contribution and the interference contribution to both the
conductance and noise.
We discuss their relation in the following.

We begin by stating the framework for the theory of current
fluctuations \cite{FCS,BN01}.
It relies on the evaluation of the quasiclassical Green's
functions $\check{g}(x,\varepsilon,\chi)$ in the
Nambu$(\,\hat{}\,)$-Keldysh$(\,\bar{}\,)$ space, at a given energy
$\varepsilon$ and counting field $\chi$.
The counting field appears as a gauge transformation of the
Green's function in the normal reservoir
\beq \label{eq:gn}
  \check g_{N}(\chi)
  =
  e^{-\frac{i}{2}\chi\check\tau_{K}}\,
  \check g_{N}^{0} \,
  e^{\frac{i}{2}\chi\check\tau_{K}}
  \quad ,
\eeq
where $\check g_{N}^{0} = \hat\tau_3\otimes\bar\sigma_3+
(f_{T0}+f_{L0}\hat\tau_3)\otimes(\bar\sigma_1+i\bar\sigma_2) $ is
the standard Green's function for a metallic reservoir,
$\bar\sigma_i,\hat\tau_{j(i,j=1,2,3)}$ are Pauli matrices,
$\check\tau_{K}=\hat\tau_3\otimes\bar\sigma_1$,
$f_{L0}=1-f_+-f_-$, $f_{T0}=f_--f_+$,
$f_{\pm}(\varepsilon)=f(\varepsilon\pm eV)$, $f$ is the Fermi
function at temperature $T$, and $V$ is the voltage bias.
Given $\cg(x)$, one can calculate the spectral matrix current in
the wire
\begin{equation}\label{eq:matrix-current}
    \check{J}(x)=-L\, G_D\, \, \check{g}(x) \partial_{x} \check{g}(x)
    \quad,
\end{equation}
with $x$ the coordinate along the wire of length $L$.
The current and the zero-frequency noise are given respectively by
\cite{BN01}
\begin{equation} \label{IandS}
    I = J(\chi=0)
    \quad \text{and} \quad
    S =
    \left.
    -2i\,e
    \frac{\partial J(\chi)} {\partial \chi} \right|_{\chi=0}
    \quad ,
\end{equation}
where $
    J(\chi)
    =
    -1/(8e)
    \int \!\! d\varepsilon\,  \text{Tr} \left[\check\tau_K \check J(x) \right]
$
and  $J(\chi)$ does not depend on the position $x$ where $\check J$
is evaluated.

The Green's function $\cg(x)$ in the wire is determined by the
diffusion-like Usadel equation
\beq\label{eq:usadel}
   -\frac{\hbar \, D}{L\,G_D}
   \partial_{x} \check{J}(x)
   =
   -i \varepsilon
   \left[\check{\tau}_3,\check{g}(x)\right] \, ,
\eeq
(where $\check \tau_3=\hat\tau_3 \otimes \bar 1$ and $D$ is the
diffusion constant in the wire) and by boundary conditions
at the two extremities.
We assume a good contact on the normal side at $x=L$.
This implies $\cg(L) = \cg_N$ as defined in \refe{eq:gn}.
On the superconducting side, at $x=0$, the boundary condition expresses the
conservation of the spectral matrix current (\ref{eq:matrix-current})
through the interface \cite{zaitsev,CircuitTheory}:
\beq
  \check{J}(0)
  =
  G_Q \sum_n
  \frac{2\myT_n\left[\check{g}(0),\check{g}_{S}\right]}{
  4+\myT_n(\{\check{g}(0),\check{g}_{S}\}-2)} \quad ,
  \label{eq:boundary}
\eeq
where the eigenvalues $\myT_n$ of the transmission matrix through
the interface appear explicitly and $G_B=G_Q \sum_n \myT_n$, with
$G_Q=e^2/(\pi\hbar)$ the quantum of conductance.
The Thouless energy, $E_{T}=\hbar D/L^2$, is the relevant energy
scale in \refE{eq:usadel}.
We restrict to the case where it is much smaller than the
superconducting gap.
With the same restrictions on the voltage bias and temperature,
the Green's function in the superconducting reservoir is
$\cg_S=\hat\tau_2 \otimes \bar 1$.
Extension to higher energies is straightforward, but it will not
be considered to keep the presentation simple.
The quasiclassical Green's functions obey the normalization
condition $\cg^2=1$ and the symmetry property
\beq \label{symmetry}
  \cg(x)^{\dag}
  =
  - \check{\tau}_{L} \, \check{g}(x) \,\check{\tau}_{L},
 \quad
 {\rm with}
 \quad
 \check{\tau}_{L}=\hat\tau_3\otimes\bar\sigma_2
 \quad .
\eeq

Finding $J(\chi)$ for all values of $\chi$ is equivalent to
calculate the full counting statistics of charge transfer
\cite{FCS} and it may be a formidable task.
One of the main difficulties comes from the normalization
condition $\cg^2=1$.
As a matter of fact, for $\chi=0$ the Green's functions have a
triangular structure in Keldysh space:
\beq
\cg_0
=
\cg(\chi=0)
=
\left(\begin{array}{cc}
  \hat{g}^{R} & \hat{g}^{K} \\
  0 & \hat{g}^{A} \\
\end{array}\right)
\quad .
\eeq
In the absence of supercurrent it exists a simple parametrization
fulfilling the normalization condition and the symmetry property
\refe{symmetry}: $\hat{g}^{R}=\hat{\tau}_{3} \cosh\theta
+i\,\hat{\tau}_{2}\sinh\theta $, $\hat{g}^{K} =
\hat{g}^{R}\hat{f}-\hat{f} \hat{g}^{A}$, and
$\hat{f}=f_{L}+\hat{\tau}_{3}f_{T}$,
with $\theta=\theta_1+i \theta_2$ and $f_L$, $f_T$, $\theta_{1}$,
and $\theta_2$ real.
When $\chi \neq 0$ the triangular structure is lost and we are not
aware of a simple parametrization for $\cg$ in this case.

A less ambitious program is to calculate only the noise.
In this case it is clear from \refE{IandS} that the full
dependence of $\cg(\chi)$ is not needed:
$\partial\cg(\chi)/\partial\chi|_{\chi=0}$ is enough.
As suggested in Ref.~\cite{BN01} one can thus
develop $\cg(\chi)$ in powers of $\chi$:
\beq
  \cg(x,\chi)=\cg_0(x)-i(\chi/2) \check g_1(x)+{\cal O}(\chi^2)
\eeq
and solve the problem order by order in $\chi$.
For the leading order Green's function, $\cg_0$, the above
parametrization is appropriate.
\refE{eq:usadel} for $\hat g^R$ gives the non-linear equation
\beq \label{eqtheta}
   \hbar\,D \, \theta''(x)+2i\varepsilon\sinh\theta(x)=0
   \,,
\eeq
with boundary conditions $\theta(L)=0$ and
\beq \label{BCtheta}
    L\, \theta'(0)
    = {i\over r}
    \qav{
    \cosh \theta(0)
    \over 1+{\myT\over 2}
    \left(i \sinh \theta(0)-1\right)
} \,.
\eeq
We defined $
  \qav{\psi (\myT)}
  \equiv
  \sum_n \myT_n \psi(\myT_n)/\sum_n \myT_n
$
and
$r=G_D/G_B$.
\refE{eq:usadel} for $\hat g^K$ gives $f_L(x)=f_{L0}$ and the
differential equation for $f_T$:
$(\cosh^2 \theta_1(x) f_T'(x))'=0$
with boundary conditions $f_T(L)=f_{T0}$ and
\beq \label{BCfT}
 f_T'(0)
 =
 {f_T(0)\, \theta'_1(0) \over \cosh \theta_1(0)\,
 \sinh \theta_1(0)}
 \quad.
\eeq
Then the current is
$
    I = 1/(2e)
    \int \!\! d\varepsilon \,{\cal G}(\varepsilon)
    f_{T0}(\varepsilon)
$
 with \cite{VZK93}
\beq \label{eqG}
    {\cal G} (\varepsilon)
    =
    G_D \left[
  {\cal D}^{-1}(\varepsilon)+
  {\tanh \theta_1(0)\over L\, \theta'_1(0)}
  \right]^{-1}
\eeq and
$
  {\cal D}^{-1}(\varepsilon)
  =
  1/L \int_0^L ds/\cosh^{2}\theta_1(s). $
At low temperatures, $k_B T\ll eV$, $G(V)={\cal G}(eV)$ is the
differential conductance.
These equations have been recently exploited in Ref.~\cite{tanaka}
to discuss the conductance.

At the next order in $\chi$, we get a linear matrix differential
equation for $\cg_1(x)$:
\beq \label{eqg1}
    \hbar \,D \, \partial_x \check{J}_1(x) =
    i\,\varepsilon\,L\,G_D\,
    \left[\check{\tau_3},\check{g}_1(x)\right]
\eeq
with $\check{J}_1(x)=-L \,G_D\, [\check{g}_0(x)\partial_{x}
\check{g}_1(x)+\check{g}_1(x)\partial_{x} \check{g}_0(x)]$. The
boundary conditions read
$\check g_1(L)=[ \check{\tau}_K, \check g_N^0]$
on the normal side and
$ \check{J}_1(0) = 2 G_B \langle \check A \check B \check A
\rangle $ where
$\check A = [4+\myT(\{\check{g}(0),\check{g}_{S}\}-2)]^{-1} $
 and
$ \check B = [4+2 \myT
    (\check{g}_{S}\check{g}_{0}(0)\check{g}_{1}(0)\check{g}_{S}-
                \check{g}_{0}(0)\check{g}_{1}(0)-[\check{g}_{1}(0),\check{g}_{S}])
] $
on the superconducting side.
Finally, the normalization of $\cg$ gives the condition
$\{\check g_0(x),\check g_1(x)\}=0$ that can be fulfilled with the
change of variable
$\check g_1(x)=[\check g_0(x),\check\phi(x)]$.
The matrix $\check \phi(x)$ is now constrained  by the symmetry
properties \refe{symmetry} only.
We find that \refE{eqg1} can be conveniently solved with the following
parametrization for $\check \phi$:
\begin{equation}\label{parametrization}
    \check\phi
    =
    \left(
    \begin{array}{cc}
      a \, f_{T0}\hat\tau_1-c\, \hat f\, \hat\tau_3 & b\, \hat\tau_3 + d \\
      c\, \hat\tau_3 &  a^*\, f_{T0}\, \hat\tau_1+c \,\hat f \, \hat\tau_3 \\
    \end{array}
    \right)
\end{equation}
with $a=a_1+i \,a_2$ and $a_1$, $a_2$, $b$, $c$, $d$ real
functions of $x$.
Substituting $\cg_1$ in terms of $\check \phi$ into \refe{eqg1} after
straightforward but lengthly calculations we obtain the
set of equations and boundary conditions for the four
parameters $a$, $b$, $c$, and $d$.
\begin{widetext}
The parameter $a$ plays the role of $\theta$ in the lowest order
equation \refe{eqtheta}:
\beq
 \label{eqfora}
 \hbar\,D \,a''(x) + 2i\, \varepsilon\,  a(x) \cosh\theta(x)
 =
 -2 E_{T}\,
 {\sinh\theta_1(x) \over \cosh^3 \theta_1(x)}
 \frac{G(\varepsilon)^2}{G_D^2}
 \,.
\eeq
The boundary conditions are $a(L)=0$ and $L\,a'(0)=\alpha\, a(0)/r
+\beta/r$ with

\beq \label{BCa}
\alpha = \left\langle { i \sinh \theta-\myT(i\sinh
    \theta-1)/2 \over \left[1+\myT(i\sinh \theta-1)/2\right]^2 }
\right\rangle
\,,
\qquad
\beta = {i c^2 \over 8} \left\langle {2\myT^2\cosh
\theta^*+8(\myT-1) \cosh\theta -2 i\myT(\myT-2)\sinh \theta \cosh
\theta^* \over \left|1+\myT(i\sinh \theta-1)/2\right|^2
\left(1+\myT(i\sinh \theta-1)/2\right) } \right\rangle \,, \eeq
\end{widetext}
both evaluated at $x=0$.
 The equation for $b$ resembles that for
$f_T$: it is solved analytically in terms of an integral that
enters the final expression for the noise.
The parameter $c$ turns out to be simply proportional to $f_T(x)$:
$c(x)=-f_T(x)/f_{T0}$ with $c(0)=1-{\cal G}(\varepsilon)/[G_D{\cal
D}(\varepsilon)]$.
Finally we find $d(x)=2 f_{L0} f_{T0}[1+c^3(x)-2\tan
\theta_2(x)]$.
Its knowledge is not necessary to obtain the noise.

The expression for the noise is obtained by evaluating
\refE{IandS}:
\begin{equation} \label{eq:bruit}
  S
  =
  \int \!\! d\varepsilon \, {\cal G}(\varepsilon)
  \left\{ 1-f_{L0}^2(\varepsilon)-[1-{\cal F}(\varepsilon)]f_{T0}^2(\varepsilon) \right\},
\end{equation}
where
\begin{eqnarray}
\lefteqn{ {\cal F}(\varepsilon)=\frac{2}{3}(1+c(0)^3)
+ {2{\cal G}(\varepsilon)\over G_D} \int_0^1 {\sinh \theta_1 a_1 \over \cosh^3
\theta_1}ds}
\nonumber \\
&&
-c(0)
\left(
     {G_D a_1'(0) c(0) \over {\cal G}(\varepsilon) \tanh \theta_1(0)}
     +{2 a_1(0) \over \sinh 2\theta_1(0)}
\right)
\,.
\label{eq:fano}
\end{eqnarray}
In equilibrium, Eq. (\ref{eq:bruit}) yields the fluctuation
dissipation relation $S=4k_B\,T\,G$, as expected.
Out of equilibrium, in the shot-noise regime $k_B\, T \ll eV$,
\refE{eq:fano} defines the experimentally accessible differential
Fano factor $F(V) \equiv (dS(V)/dV)/(2e G(V)) = {\cal F}(eV)$.

\refE{eq:fano} is the central result of the paper.
Once the problem for the conductance has been solved and
$\theta(x)$ and $f_T(x)$ are known \cite{VZK93,tanaka}, we provide
a simple and efficient way to calculate the noise.
It suffices to solve the linear differential equation
\refe{eqfora} with the given boundary conditions, and substitute
the result into \refE{eq:fano}.
This program has to be followed numerically in most situations,
but its implementation is straightforward and allows to obtain
quantitative predictions for a wide range of realizable
experiments.
The previous approach to the same problem has been to discretize
\refE{eq:usadel} from the outset and solve it numerically
\cite{CircuitTheory,BN01}.
That technique is particularly efficient to treat the simplified boundary
conditions
$\check{J}(0) = 2 G_B \left[\check{g}(0),\check{g}_{S}\right]$ \cite{KL},
valid only when all $\myT_n\ll 1$.
Its application to the general case is numerically much more difficult.

We can now discuss the crossover from the coherent
($\varepsilon\ll E_{T}$) to the incoherent semiclassical
($\varepsilon\gg E_{T}$) regime quite generally.

In the fully coherent regime we recover known expressions obtained
with random-matrix theory \cite{dJB94,Beenakker97} or
quasiclassical Green's functions \cite{Nazarov94}.
In the opposite limit, propagation is incoherent: due to the large
value of $\varepsilon$ the phases accumulated along two different
paths are always uncorrelated.
In order to verify this point we first repeated the calculation
when both reservoirs are normal.
It suffices to substitute $\cg_S$ with $\cg_N$ into
\refe{eq:boundary} with $eV=0$.
(The procedure greatly simplifies since $\cg$ becomes diagonal in
Nambu indexes.)
As expected, we find no energy dependence for ${ G}$ and ${ F}$.
For the conductance the usual Ohm's law holds:
$G^{-1}=G_D^{-1}+G_B^{-1}$ while for the Fano factor we find
\beq
\label{FanoNormal}
  F
  =
  \frac{1}{3}
  \left[
    1+
    \left(
       2-3 {\sum_n \myT_n^2 \over \sum_n \myT_n }
    \right) \frac{G_D^3}{(G_D+G_B)^3}
  \right]
  \,.
\eeq
\refE{FanoNormal} coincides with the semiclassical result obtained
in Ref.\ \cite{deJongBen}.
In the superconducting case, \refE{eqG} and \refE{eq:fano} can be
evaluated analytically for $\varepsilon \gg E_{T}$ defining
$G_{\inc}$ and $F_{\inc}$, respectively.
It turns out that the conductance and the Fano factor are given by
the same expressions for the normal case multiplied by a factor
$2$ and where the substitutions $G_D\rightarrow G_D/2$ and $\myT_n
\rightarrow \myT_n^A=\myT_n^2/(2-\myT_n)^2$ have been operated
(the last one implies $G_B\rightarrow G_B \qav{\myT/(2-\myT)^2}$).
This result can be understood \cite{Bezuglyi,SB,BelzigSamuelsson}
in terms of semiclassical
probability for charge transfer, that is the probability for an
electron to cross the distance $L$ to the barrier, to be Andreev
reflected, and then to come back as a hole on the same distance.
Therefore, the wire in contact with the superconductor can be
described by two normal wires, one for the electrons and the other
for the holes, coupled through a fictitious barrier that converts
electrons into holes with probability $\myT_n^A$ and to which a
voltage bias of $2\,V$ is applied \cite{Bezuglyi}.

Quantum corrections to both $G(\varepsilon)$ and $F(\varepsilon)$
can be evaluated as an expansion for large $\varepsilon$.
We give the result for all $\myT_n=\myT$:
$G=G_{\inc}+G_1(\varepsilon) $
, with
$G_1
 =
 2\,\sqrt{E_{T}/\varepsilon}\,( 4 - 4\myT- \myT^2)/( r\, (\myT-2)^2 + 2\,\myT )^2
$, and
\beq \label{Fpheno}
    F(\varepsilon) = F_{\inc}-r\,\gamma(\myT,r) [G(\varepsilon)-G_{\inc}]
\eeq
with
$ \gamma
   =
   [
    r( 2 - \myT )^2
     (
        64 + ( 2 - \myT ) \,\myT\,
        (
            64 + r\,( 2 - \myT ) \,(  12-12\myT -\myT^2 )
        )
     )
   ]
   /
   [
     ( r( 2- \myT ) ^2 + 2\,\myT)^2\, ( 4 -4\myT- \myT^2)
   ]
$.
$\gamma$ is positive for most values of $\myT$.
\refE{Fpheno} relates the interference contributions to $G$ with
those to $F$ for large $\varepsilon$.
When the barrier dominates ($r \gg 1$) the relation \refe{Fpheno}
holds for any value of $\varepsilon$ up to first order in $1/r$
with $\gamma(\myT) = \gamma(\myT, r \rightarrow \infty)$ and
$ G_1(\varepsilon) = 2\,\phi(2\sqrt{\varepsilon/E_{T}}) \,( 4
-4\myT- \myT^2)/ [r^2\, (\myT-2)^4] $
 and $\phi(x)=(\sinh x + \sin
x)/[x(\cosh x + \cos x)]$.
This suggests that the simple relation \refe{Fpheno} stands,
though approximately, beyond the range of validity for which it
has been proved.
A possible interpretation is that interference modifies the
effective transparency of the whole system.
The linear relation between the quantum corrections to $G$ and $F$
then corresponds to the single channel quantum result: $G\propto
\myT$ and $F\propto 1-\myT$ \cite{BlanterButtiker}.

For intermediate values of the parameters $r$, $\varepsilon$, and
$\myT_n$, quantum interference contribution can be studied
numerically.
We report the results in Fig.\ 1 for $G_D=G_B$ and different
values of $\myT=\myT_n$ for all $n$.
\begin{figure}
\centerline{\psfig{file=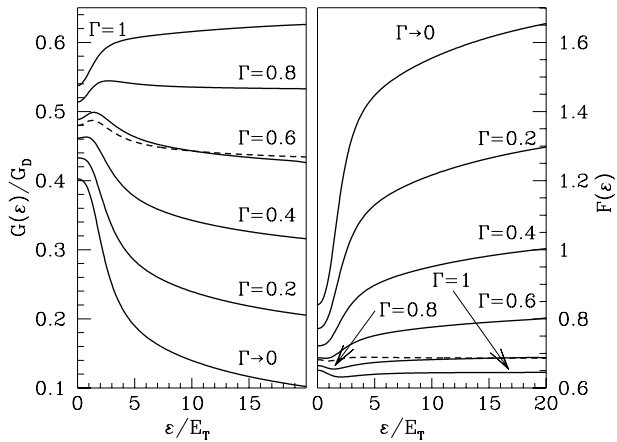,angle=-90,width=8.3cm}}
\caption{
Energy dependence of the conductance and of the differential
Fano factor when the wire and the interface have the same
conductance ($G_D=G_B$) for different values of $\Gamma=\Gamma_n$.
The dashed lines correspond to the universal distribution $\{\myT_n\}$
for disordered interfaces.
}
\end{figure}
The transparency $\myT$ drives a smooth crossover from a
reflectionless tunneling to a reentrant behavior which can be seen
both in the conductance and the Fano factor.
This proves that the energy dependence of both may be strongly
affected by the precise set $\{ \myT_n \}$ of transparencies of
the interface.
As a relevant example we plot the result for the
universal distribution for disordered interfaces
$\sum_n\delta(\myT-\myT_n)\propto 1/(\myT^{3 \over
2}\sqrt{1-\myT})$ \cite{SchepBauer}.
Fig. 1 also confirms that \refE{Fpheno} is actually qualitatively
satisfied in the whole range of parameters.

In the limit $G_D \ll G_B$ ($r \rightarrow 0$) the actual values
of $\{ \myT_n \}$ drop from the boundary conditions.
Indeed Eqs.~(\ref{BCtheta},\ref{BCfT},\ref{BCa}) force
$\theta(0)=-i\pi/2$, $f_T(0)=0$ (and thus $c(0)=0$), and $a(0)=0$.
The conductance becomes ${\cal G}(\varepsilon)={\cal
D}(\varepsilon) G_D$ and the Fano factor is given by
\refE{eq:fano} where only the $2/3$ and the integral terms
survive.
The first term is the semiclassical incoherent value
\cite{NagaevButtiker}, which also coincides with the fully
coherent one \cite{Beenakker97}.
The integral term singles out the interference contribution to the
Fano factor.
The energy dependence coincides with that obtained previously in a
different way \cite{BN01} which, in turn, agrees qualitatively
with the experimental result \cite{KSP}.
However, a broader voltage range for the reentrance in $G(V)$ and
$S(V)$ is predicted \cite{impr}.
A possible explanation is that the barrier resistance is not
negligible.
Assuming a disordered interface, we find that the reasonably small
value of $r=0.3$ allows to fit the conductance.
It also improves the fit for the noise, but the agreement is not
perfect.
This may be due either to a different distribution of
transparencies at the barrier, or, as suggested in Ref.\ \cite{BN01}, to
heating or interaction effects.

In conclusion, we provide a framework to calculate both conductance
and noise in a normal metallic wire connected to a superconducting
lead through an arbitrary interface.
We predict a strong energy dependence of the coherent contribution
to the Fano factor.
We suggest to exploit this dependence to experimentally
characterize the transparency of interfaces.

We thank V.~P.\ Mineev, M.\ Sanquer, and F.~W.~J.\ Hekking for useful
discussion.
F.P. acknowledges financial support from CNRS through contract
ATIP-JC 2002.

\end{document}